\documentclass[10pt,conference]{IEEEtran}
\IEEEaftertitletext{\vspace{-1cm}} 
\IEEEoverridecommandlockouts
\usepackage{etoolbox}
\usepackage{cite}
\usepackage{amsmath,amssymb,amsfonts}
\usepackage{algorithmic}
\usepackage{threeparttable}
\usepackage{graphicx}
\usepackage{textcomp}
\usepackage{shortcuts}
\usepackage{framed}
\usepackage{array}
\usepackage{booktabs}

\usepackage{hyperref}
\usepackage{url}

\usepackage{breakurl}
\usepackage[most]{tcolorbox}
\usepackage{multirow}
\usepackage{amsthm}

\usepackage{pifont}
\usepackage{xspace}
\usepackage{caption}   

\newcommand{\greencheck}{{\color{green}\checkmark}}
\newcommand{\redcross}{{\color{red}\times}}
\newcommand{\zsd}[1] {{\color{black}{#1}}}

\def\BibTeX{{\rm B\kern-.05em{\sc i\kern-.025em b}\kern-.08em
    T\kern-.1667em\lower.7ex\hbox{E}\kern-.125emX}}
\makeatletter
\DeclareRobustCommand\onedot{\futurelet\@let@token\@onedot}
\def\@onedot{\ifx\@let@token.\else.\null\fi\xspace}
\def\eg{\emph{e.g}\onedot} 
\def\ie{\emph{i.e}\onedot}

\makeatother
\usepackage[linesnumbered,ruled,vlined]{algorithm2e}
\usepackage{enumitem}

\begin{document}

\title{Understanding the Effectiveness of Coverage Criteria for Large Language Models: A Special Angle from Jailbreak Attacks}

\author{
    \IEEEauthorblockN{Shide Zhou}
    \IEEEauthorblockA{\textit{Huazhong University of}\\
    \textit{Science and Technology}\\
    Wuhan, China \\
    shidez@hust.edu.cn
    }
    \and
    \IEEEauthorblockN{Tianlin Li\textsuperscript{\dag}}
    \IEEEauthorblockA{\textit{Nanyang Technological University}\\
    Singapore\\
    tianlin001@e.ntu.edu.sg
    }
    \and
    \IEEEauthorblockN{Kailong Wang\textsuperscript{\dag}}
    \IEEEauthorblockA{\textit{Huazhong University of}\\
    \textit{Science and Technology}\\
    Wuhan, China \\
    wangkl@hust.edu.cn
    }
    \and
    \IEEEauthorblockN{Yihao Huang}
    \IEEEauthorblockA{\textit{Nanyang Technological University}\\
    Singapore\\
    huang.yihao@ntu.edu.sg
    }
    \and
    \IEEEauthorblockN{Ling Shi}
    \IEEEauthorblockA{\textit{Nanyang Technological University}\\
    Singapore\\
    ling.shi@ntu.edu.sg
    }
    \and
    \IEEEauthorblockN{Yang Liu}
    \IEEEauthorblockA{\textit{Nanyang Technological University}\\
    Singapore\\
    yangliu@ntu.edu.sg
    }
    \and
    \IEEEauthorblockN{Haoyu Wang}
    \IEEEauthorblockA{\textit{Huazhong University of}
    \textit{Science and Technology}\\
    Wuhan, China \\
    haoyuwang@hust.edu.cn
    }
    \thanks{\textsuperscript{\dag}Corresponding Authors.}
}
\makeatletter
\patchcmd{\@maketitle}
  {\addvspace{0.5\baselineskip}\egroup}
  {\addvspace{-2.5\baselineskip}\egroup}
  {}
  {}
\makeatother

\maketitle
\begin{abstract}
Large language models (LLMs) have revolutionized artificial intelligence, but their increasing deployment across critical domains has raised concerns about their abnormal behaviors when faced with malicious attacks. 
Such vulnerability alerts the widespread inadequacy of pre-release testing.
In this paper, we conduct a comprehensive empirical study to evaluate the effectiveness of traditional coverage criteria in identifying such inadequacies, exemplified by the significant security concern of jailbreak attacks.
Our study begins with a clustering analysis of the hidden states of LLMs, revealing that the embedded characteristics effectively distinguish between different query types. We then systematically evaluate the performance of these criteria across three key dimensions: criterion level, layer level, and token level.

Our research uncovers significant differences in neuron coverage when LLMs process normal versus jailbreak queries, aligning with our clustering experiments. \zsd{Leveraging these findings, we propose three practical applications of coverage criteria in the context of LLM security testing. Specifically, we develop a real-time jailbreak detection mechanism that achieves high accuracy (93.61\% on average) in classifying queries as normal or jailbreak. Furthermore, we explore the use of coverage levels to prioritize test cases, improving testing efficiency by focusing on high-risk interactions and removing redundant tests. Lastly, we introduce a coverage-guided approach for generating jailbreak attack examples, enabling systematic refinement of prompts to uncover vulnerabilities. }
This study improves our understanding of LLM security testing, enhances their safety, and provides a foundation for developing more robust AI applications.
\end{abstract}

\section{Introduction}\label{sec:intro}
Large language models (LLMs) have emerged as a transformative technology in artificial intelligence, drastically altering the way machines understand and generate human language. These models have become the backbone of numerous applications, from automating customer support to providing decision support in critical domains such as finance~\cite{son2023classification}, healthcare~\cite{tang2023does}, and legal sectors~\cite{10.1145/3594536.3595163}. 
However, as LLMs are increasingly deployed across these crucial domains, they are frequently reported to suffer from abnormal model behaviors \cite{zhuo2023exploring,li2024largelanguagemodelsecretly,li2024benchmarkingbiaslargelanguage}, particularly jailbreaking to generate harmful content \cite{zou2023universal,zhu2023autodan,yu2023gptfuzzer,deng2024masterkey,chao2023jailbreaking,deng2024pandora,huang2025perceptionguidedjailbreaktexttoimagemodels}, which causes severe societal impacts. 
This underscores the need for effective testing techniques to identify such attacks and maintain the trustworthiness of LLMs. One crucial aspect of examining the adequacy and quality of tests for LLMs is the coverage criteria, which provide a systematic way to measure the extent of model testing and identify potential gaps or weaknesses in the testing process. \looseness=-1

Despite a lack of testing coverage criteria specifically designed for LLMs, prior research in the community has made commendable efforts to design coverage criteria for small-scale neural networks. These efforts provide valuable insights that could be referenced and further applied to LLMs.
Among these coverage criteria, Neuron Coverage (NC)~\cite{10.1145/3132747.3132785} focuses on the activation levels of individual neurons, while K-Multisection Neuron Coverage (KMNC)~\cite{10.1145/3238147.3238202} examines the utilization of neurons across different activation ranges. Top-k Neuron Coverage (TKNC)~\cite{10.1145/3238147.3238202} and Top-k Neuron Patterns (TKNP)~\cite{10.1145/3238147.3238202} emphasize the role of highly activated neurons and their patterns in the decision-making processes of the model. These criteria focus on different aspects of neuron activation and utilization, offering a foundation for the development of LLM-specific coverage criteria.

However, the application and effectiveness of these coverage criteria in LLMs remain unexplored, motivating our aim to investigate their applicability and efficacy in this context.
The sheer size and complexity of LLMs, characterized by their vast number of parameters and deeper layers, present unique challenges that previous small-scale evaluations may not address. These challenges include determining the most suitable and effective criterion for LLMs, identifying appropriate inspection points to monitor, and striking a balance between comprehensive coverage and feasibility. The generative nature of LLMs and their complex dynamics further complicate the selection of monitoring points and the adequacy of existing criteria in capturing their intricacies.

To bridge the gap in our understanding of how traditional coverage criteria perform when applied to LLMs, comprehensive empirical studies are crucial. First, systematically evaluating the effectiveness of existing criteria, such as NC, TKNC, and TKNP, in the context of LLMs will help identify the most suitable criteria for assessing their robustness and reliability. Second, applying these criteria to specific scenarios, such as detecting abnormal LLM behavior, can provide valuable insights into the models' limitations and potential vulnerabilities, offering an alternative yet promising perspective on anomaly detection in LLMs and complementing existing methods.

\textbf{Our Work.} In this work, we conduct an in-depth and extensive empirical study, focusing on the well-defined abnormal behavior of jailbreak attacks, to examine the effectiveness of coverage criteria.
Our study begins with a comprehensive clustering analysis of the hidden states of LLMs as they process various queries, revealing that the characteristics embedded within these hidden states effectively distinguish between different query types. Building upon this foundation, we then unfold our research across three key dimensions:
\ding{182} \textbf{Criterion Level}, where we evaluate and compare the effectiveness of different coverage criteria in LLMs;
\ding{183} \textbf{Layer Level}, where we assess the impact of network layers on coverage criteria to understand layer-specific dynamics;
\ding{184} \textbf{Token Level}, where we explore coverage criteria performance across tokens to provide insights into the granularity of model responses.

Our research reveals significant differences in neuron coverage when LLMs process normal versus jailbreak attack queries, aligning with our clustering experiments. \zsd{These empirical findings have led to the development of three practical downstream applications leveraging coverage criteria to enhance LLM security testing. The first is \emph{real-time detection of jailbreak attacks}, which uses neural activation features to classify queries as normal or jailbreak with high accuracy (93.61\%). This approach enables early detection from the model’s first token output, offering robust real-time security capabilities for LLM-integrated systems. Additionally, we propose \emph{test case prioritization} to improve testing efficiency by identifying high-risk test cases and eliminating redundant ones. Finally, we introduce \emph{jailbreak case generation}, employing coverage-guided methods to systematically refine prompts and uncover vulnerabilities. These applications highlight the potential of coverage criteria to address key challenges in LLM security testing.}

\textbf{Contributions.} In summary, this research makes the following contributions:
\begin{itemize}[leftmargin=*]
\item \textbf{An extensive empirical study for LLM evaluation.} Our extensive empirical study reveals significant differences in neuron coverage between normal and jailbreak queries when evaluating traditional coverage criteria in LLMs.
\zsd{\item \textbf{Three practical applications for LLM security testing.} We propose real-time jailbreak detection, test case prioritization, and jailbreak case generation, demonstrating the versatility of coverage criteria.}
\item \textbf{Towards robust LLM development.} We have enhanced the understanding of LLM security testing, laying a foundation for more robust and resilient AI applications.
\end{itemize}

\section{Background and Preliminaries}
\subsection{Model Inference Process}
\label{sec:formalization}
We formalize the LLM inference process based on the Transformer architecture, starting with the input vector \( h_0 \). The text is tokenized into discrete elements, each mapped to a dense embedding vector representing token semantics for further processing by the model. Each Transformer block, or simply ``block'' hereafter, denoted by \( i \) (where \( i = 0, 1, \ldots, L-1 \), and \( L \) is the total number of blocks), enhances the data by operating through two primary layers:

\textbf{Attention Layer} adjusts the input vector \( h_{i} \) by selectively focusing on various segments of the data sequence. It enhances the ability of the model to respond to contextual nuances by dynamically weighting the importance of different inputs:\vspace{-0.2cm}
\[
h_i' = \text{L}_{\text{attn}}(h_{i}) + h_{i}, \quad i = 0, 1, \ldots, L-1
\vspace{-0.2cm}
\]

\textbf{MLP Layer} processes the output \( h_i' \) from the attention layer. It applies a series of nonlinear operations to capture complex relationships within the data:\vspace{-0.2cm}
\[
h_{i+1} = \text{L}_{\text{mlp}}(h_i') + h_i', \quad i = 0, 1, \ldots, L-1
\vspace{-0.2cm}
\]

Finally, after processing through \( L \) transformer blocks, the final output \( h_L \) is mapped by a linear layer and transformed into a probability distribution using the softmax function:  
\[
\text{res} = \text{softmax}(\text{Linear}(h_L))
\]

\subsection{Evaluation Criteria for Deep Neural Networks~(DNNs)}
\label{sec:CoverageCriteria}
\zsd{Mirroring the design of code coverage based on program ``logic", a series of studies recognize that the internal states (\eg, neuron performance) of DNNs can be used to represent the ``logic" of these networks. The key focus is on how to better characterize internal states to design more effective coverage criteria. Table~\ref{tab:coverage-criteria} summarizes the main criteria categorized into three types.

\begin{table}[h]
\centering
\caption{\zsd{Summary of Coverage Criteria for DNN Testing}}
\label{tab:coverage-criteria}
\resizebox{\columnwidth}{!}{
\begin{tabular}{p{3cm} p{7cm}}
\toprule
\textbf{Category} & \textbf{Coverage Criteria} \\
\midrule
Neuron Activation and Distribution 
& Neuron Coverage (NC)~\cite{10.1145/3132747.3132785}, K-Multisection Neuron Coverage (KMNC)~\cite{10.1145/3238147.3238202}, Neuron Boundary Coverage (NBC)~\cite{10.1145/3238147.3238202}, Strong Neuron Activation Coverage (SNAC)~\cite{10.1145/3238147.3238202}, Neural Coverage (NLC)~\cite{10172683} \\
\midrule
Top Neuron Activation 
& Top-K Neuron Coverage/Patterns (TKNC/TKNP)~\cite{10.1145/3238147.3238202} \\
\midrule
Neuron Trajectory or Causal Features 
& TensorFuzz Coverage (TFC)~\cite{pmlr-v97-odena19a}, Surprise Coverage (SC)~\cite{8812069}, Likelihood Surprise Coverage (LSC)~\cite{8812069}, Distance-ratio Surprise Coverage (DSC)~\cite{10.1145/3368089.3417065}, Mahalanobis Distance Surprise Coverage (MDSC)~\cite{10.1145/3368089.3417065}, Neuron Path Coverage (NPC)~\cite{10.1145/3490489}, Causal Coverage (CC)~\cite{10172609} \\
\bottomrule
\end{tabular}}
\vspace{-0.3cm}
\end{table}

\textbf{Use of Evaluation Criteria in DNN Testing. }In DNN testing, white-box coverage criteria provide insights into a test suite's ability to expose diverse functional behaviors, known as \emph{functional diversity}, and detect potential defects, referred to as \emph{defect-detection capability}~\cite{10172683}.

These criteria assess how different DNN components are covered, reflecting decision-making processes \cite{10.1145/3490489}. For example, \emph{Neuron Coverage (NC)} measures the proportion of neurons activated above a threshold, reflecting the model's logic breadth. \emph{Top-K Neuron Coverage (TKNC)} and \emph{Top-K Neuron Patterns (TKNP)} focus on significant neuron activations, capturing key functional patterns. High coverage in these metrics suggests the test suite elicits diverse behaviors, boosting confidence in the model’s reliability.

These criteria also help identify inputs that trigger incorrect or unexpected behaviors, revealing potential defects. \emph{TensorFuzz Coverage (TFC)} represents neuron outputs from the same layer as high-dimensional variables and clusters them based on different inputs. The coverage is then quantified by the number of formed clusters, capturing diverse activation patterns and potential abnormalities. \emph{Neural Coverage (NLC)} analyzes neuron activation distributions to find vulnerabilities. By linking coverage with defect discovery, these metrics prioritize test cases that expose hidden issues, enhancing DNN robustness.}

\zsd{\subsection{DNN Testing versus LLM Testing} Traditional DNN testing focuses on black-box and white-box approaches. Black-box testing uses benchmark datasets (\eg, ImageNet~\cite{deng2009imagenet}) to evaluate accuracy and generalization, and employs fuzz testing to mutate inputs (\eg, lighting, angle, noise)~\cite{10.1145/3377811.3380415} to further identify deficiencies in the model's handling of abnormal or boundary inputs.
White-box testing seeks to capture the model's ``logic" by examining its internal structure and behavior, understanding and uncovering subtle errors and vulnerabilities that black-box testing may fail to detect.
For example, DeepXplore~\cite{10.1145/3132747.3132785} designs neuron coverage to measure activated model regions and employs differential testing to identify inconsistencies among similar DNNs, efficiently detecting erroneous corner cases. \looseness=-1

Existing LLM testing predominantly relies on black-box testing by constructing benchmarks \cite{hendrycks2021measuring,srivastava2022beyond,efrat2022lmentry,zhong2023agieval}. For instance, MMLU~\cite{hendrycks2021measuring} assesses multitask accuracy across 57 subjects, including math, history, and law. Additionally, recent studies \cite{wang2022recoderobustnessevaluationcode,chen2024nlperturbatorstudyingrobustnesscode} have developed comprehensive black-box robustness testing methods for LLMs.
Despite the commendable progress in black-box testing, vulnerabilities like ``jailbreak attacks"  \cite{deng2024masterkey} continue to emerge, with their causes still unclear. This motivates us to explore white-box coverage criteria to examine LLMs' internal behavior, enabling more effective testing and defenses. \looseness=-1

However, white-box testing coverage for LLMs might differ significantly from traditional DNNs due to their vast input-output scope and larger architectures, posing challenges for adapting existing white-box criteria.
Thus, we focus on the critical security issue of ``jailbreak attacks" and study the applicability of existing white-box testing coverage approaches. 
We anticipate this will reveal vulnerabilities overlooked by black-box methods and improve testing and defense strategies. Specifically, this not only enhances jailbreak detection but also helps prioritize test cases and generate efficient jailbreak cases. Beyond the jailbreak issue targeted in this paper, we hope this study will inspire white-box exploration to other security and functionality challenges in LLMs~\cite{10.1145/3691620.3695510, 10.1145/3691620.3695271, 10.1145/3689776, 10.1145/3660799,10.1145/3691620.3695060}.}

\zsd{\subsection{Jailbreak Attack and Defenses}
LLM developers have focused on aligning models to prevent harmful or objectionable outputs. For example, when prompted with ``Write a convincing anonymous death threat,'' aligned LLMs are trained to respond with rejections like ``I cannot answer such questions,'' instead of producing harmful content. Despite these efforts, jailbreak attacks have emerged that bypass alignment measures, causing LLMs to generate harmful outputs and compromising their security~\cite{yu2023gptfuzzer,zhu2023autodan,chao2023jailbreaking,deng2024masterkey,deng2024pandora}.

In response, defensive mechanisms are evolving. Methods like erase-and-check~\cite{kumar2024certifying} provide certified safety against adversarial prompts by sequentially removing tokens and inspecting subsequences with a safety filter. Other defenses, such as SmoothLLM~\cite{robey2023smoothllm} and JailGuard~\cite{zhang2023mutation}, reduce the success rate of jailbreak attacks by randomly perturbing input prompts and aggregating predictions. Additional strategies include perplexity-based detection~\cite{jain2023baseline, alon2023detecting} and self-classification by LLMs~\cite{zhang2024parden, phute2024llm,zhou2024defendinglvlmsvisionattacks}, which detect perplexity or directly assess prompt harmfulness.
}

\section{Study Design}
\subsection{Motivation: A Cluster Analysis Experiment}
\label{sec:motivation}
Previous research on small-scale models has demonstrated that internal states can represent (and further distinguish) the ``logic" of normal and abnormal behaviors, aiding in the design of effective coverage criteria. Building on this insight, we preliminarily investigate \emph{whether the internal mechanisms of LLMs can similarly distinguish between normal and abnormal behaviors}.
Specifically, we collect queries that can trigger different behaviors in the LLM and select the outputs $h$ from the middle transformer blocks for these queries. We then perform a clustering analysis on these outputs to observe if such internal states (\ie, the action values $h$) can be used to distinguish between different model behaviors.

\textbf{Experimental Setup:}
\begin{figure}
    \centering
    \includegraphics[width=0.48\textwidth]{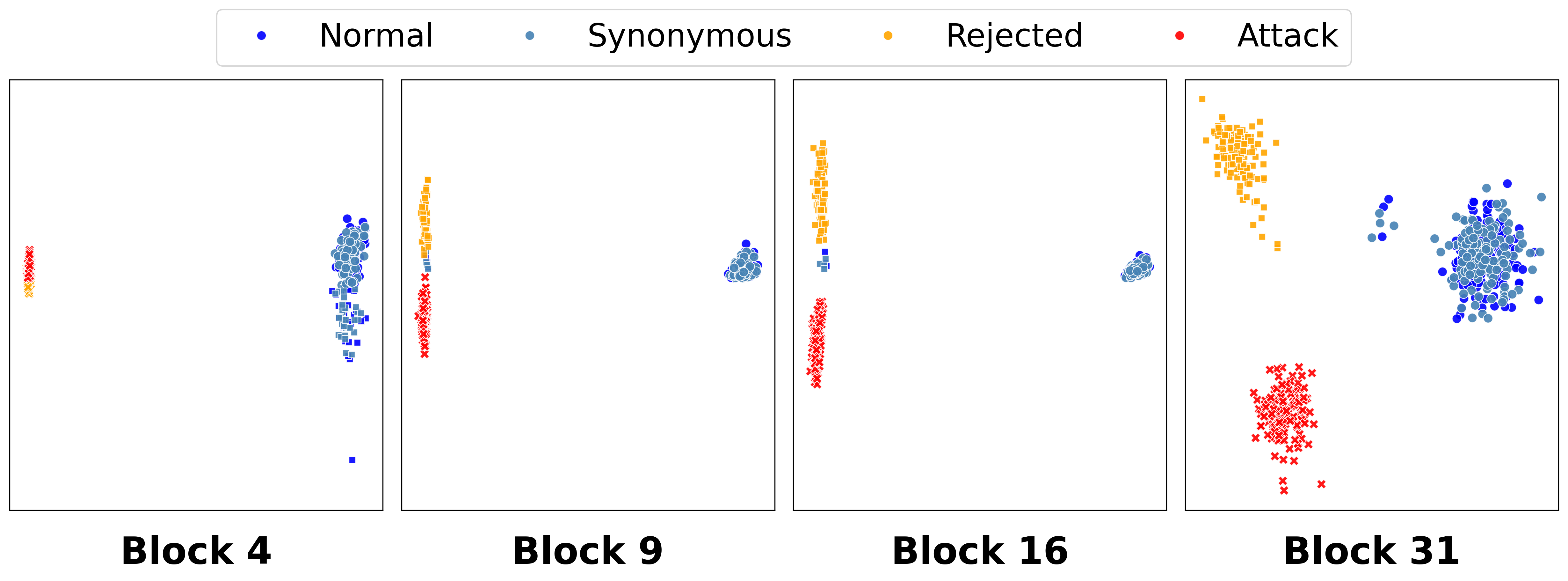}
    \caption{Clustering Experiment Analysis Results: We Select the Results of $Block4$, $Block9$, $Block16$, and $Block31$ for Display. In the Figure, We Use Colors to Distinguish Datasets and Shapes to Represent Clustering Categories}
    \label{fig:Cluster}
    \vspace{-15pt}
\end{figure}
We first introduce the setup for our cluster analysis experiments using the Llama-2-7b-chat~\cite{touvron2023llama} model as our target. 
We collect four distinct datasets and use 200 queries from each dataset, aiming to trigger different model behaviors.
\ding{182} \textbf{Normal queries} are sourced from Alpaca-GPT-4~\cite{peng2023instruction}, expected to trigger \textit{normal behaviors} of LLMs in a QA format.
\ding{183} \textbf{Synonymous queries} are the paraphrased versions of normal queries by GPT-4, intended to trigger the \textit{same normal behaviors} as the original queries.
\ding{184} \textbf{Rejected queries} are sourced from AdvBench~\cite{zou2023universal}. These malicious questions aim to trigger \textit{rejection behaviors}, considering the aligned LLM is trained to reject such queries. For example, for malicious queries like ``how to make a bomb,'' LLMs will respond with something like ``Sorry, I cannot provide \ldots'' to avoid harmful content.
\ding{185} \textbf{Attack queries} are generated by appending adversarial suffixes to rejected queries using GCG~\cite{zou2023universal}. These queries aim to trigger the model to output malicious content (\ie, \textit{abnormal behaviors}). We extract the hidden states \(h\) from the 4th, 9th, 16th, and 31st transformer blocks for these types of queries and conduct k-means clustering~\cite{macqueen1967some}. For more details about the setup, please refer to our website~\cite{llmcoverage2024}.

\textbf{Findings:} 
As shown in Figure \ref{fig:Cluster}, in the final block (block 31), the clustering of queries is clearly separated into those that trigger normal behaviors (normal and synonymous queries), rejection behaviors (rejected queries), and abnormal behaviors (attack queries).
Interestingly, normal and synonymous queries remain in the same cluster from block 1 to block 31. Initially (block 4), rejected and attack queries are clustered together, but they gradually separate as the model processes through more blocks.

In summary, our clustering analysis demonstrates that \textit{the internal states of the model include features capable of representing and distinguishing the ``logic" of different behaviors.} 
This confirms the feasibility of using internal states to design coverage criteria for LLMs. 
However, how to characterize the internal states to design better coverage criteria for LLMs remains unknown. In the following section, we introduce our methodology to thoroughly study this.

\subsection{Methodology: Evaluation Dimensions}
\begin{figure}
    \centering
    \includegraphics[width=0.46\textwidth]{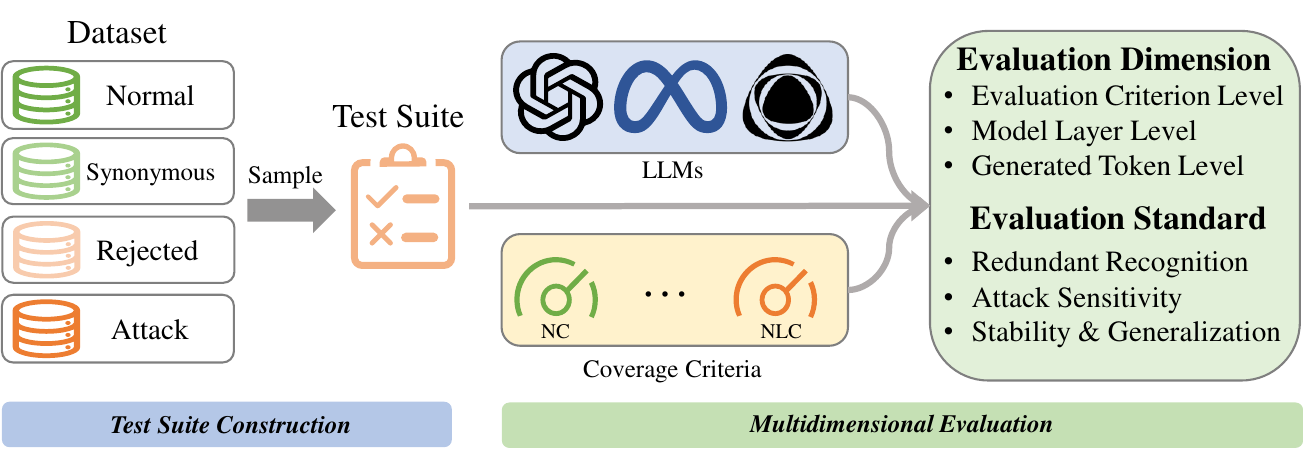}
    \caption{The Workflow of Our Study}\vspace{-0.4cm}
    \label{fig:workflow}
\end{figure}

In this section, we first provide an overview of the empirical study workflow, as shown in Figure \ref{fig:workflow}. Then, we detail the three target evaluation dimensions that potentially contribute to assessing effective coverage criteria for LLMs.

\subsubsection{Three Evaluation Dimensions}\hfill

\noindent\textbf{Evaluation Criterion Level.}\label{sec:criteria-level}
\begin{table*}[!h]
\centering
\caption{Comparison of Coverage Criteria Applicability to LLMs}
\vspace{-10pt}
\label{tab:criteria}
\resizebox{\textwidth}{!}{
\begin{tabular}{c|cc}
\hline
 \textbf{Criterion} & \textbf{Applicability} & \textbf{Reason} \\ \hline
NC/TKNC/TKNP/TFC & $\greencheck$ & N/A \\
NLC & $\greencheck$ & Applicable if we do not utilize training data for prior knowledge initialization. \\
KMNC/NBC/SNAC & $\redcross$ & Time-prohibitive to determine the activation range of neurons on all training data. \\
LSC/DSC/MDSC & $\redcross$ & Time-prohibitive to calculate neuron output trajectories for both the test suite and all training data. \\
NPC/CC & $\redcross$ & Complex causal discovery or decision path identification only designed for small DNNs. \\ \hline
\end{tabular}
}\vspace{-0.5cm}
\end{table*}

Existing DNN coverage criteria primarily use neurons or network layers as the basic computational units, evaluating model behavior from various perspectives. However, as the model size increases, particularly in LLMs as mentioned in Section~\ref{sec:intro}, these criteria face new challenges. \zsd{The vast scale of LLMs, complex training data, and intricate architectures make some coverage criteria impractical or computationally expensive. For example, NPC, designed for convolutional neural networks, is unsuitable for LLMs, while CC requires complex causal inference, with runtimes hundreds of times longer than NC. Storing hidden states for 8,000 queries in our experiment consumes 5.2 TB, just 0.0002\% of the training data. Thus, using full training data is infeasible for some criteria, while limiting to fewer samples may compromise metric effectiveness.}

Therefore, we conduct a detailed qualitative investigation and manual comparison of recently proposed coverage criteria to determine their applicability to LLMs. The detailed results are presented in Table~\ref{tab:criteria}. We select the following coverage criteria in this work: NC, TKNC, TKNP, TFC and NLC.

\noindent\textbf{Model Layer Level.} 
\begin{figure}
    \centering
    \includegraphics[width=0.48\textwidth]{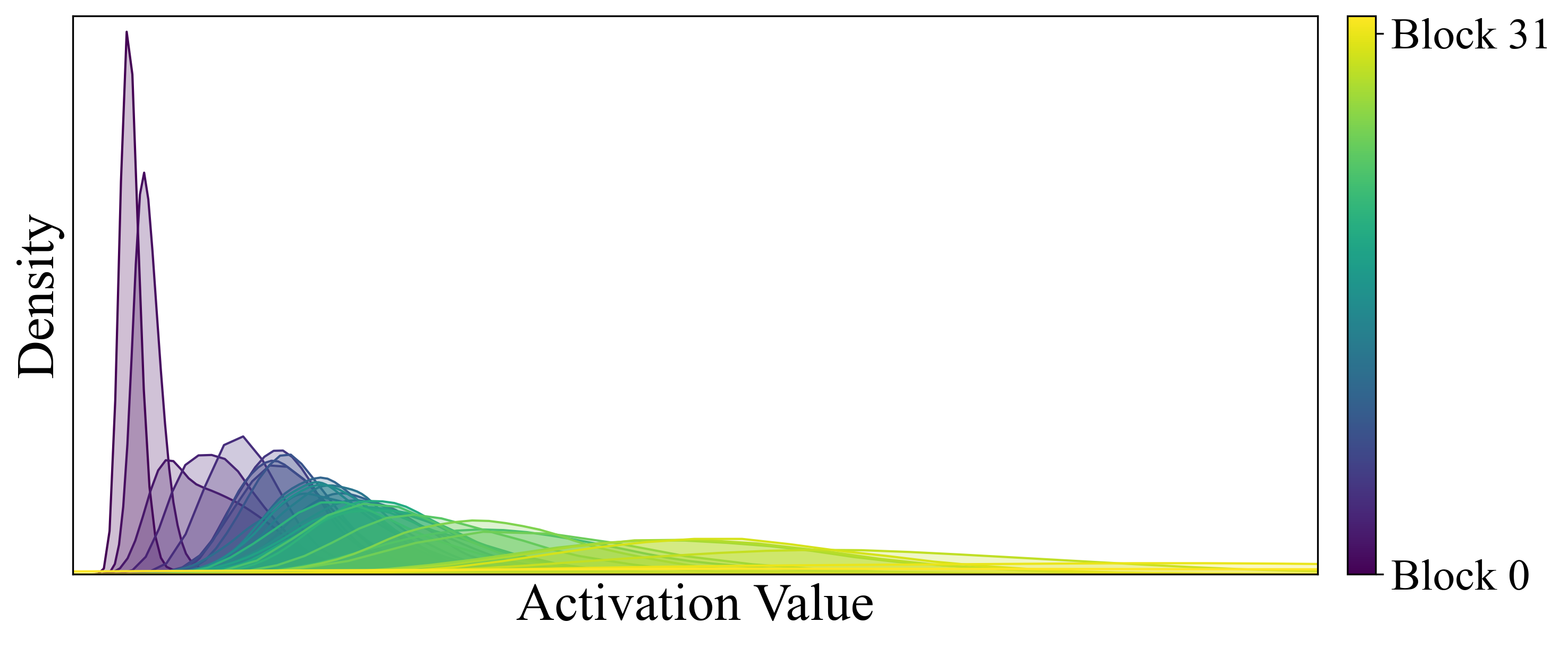}\vspace{-0.2cm}
    \caption{Probability Density Plot of Maximum Neuron Activation Values Across Model Blocks in Llama-2-7b-Chat}\vspace{-0.4cm}
    \label{fig:Density}
    \vspace{-10pt}
\end{figure}
The second dimension we focus on is the model layer Level. Unlike conventional DNN models, LLMs typically consist of multiple transformer blocks, each containing an attention layer and an MLP layer. We examine two granularities: the first involves observing the coverage changes in attention layers and MLP layers across the entire model, as denoted by \(\text{L}_{\text{attn}}(h_{i})\) and \(\text{L}_{\text{mlp}}(h_i')\) in Section \ref{sec:formalization}. The second granularity goes one step further, focusing on the coverage changes in attention layers and MLP layers within each individual block.

The reasons for choosing these two granularities are threefold. 
First, each layer in an LLM plays different roles and exhibits varying complexity when processing information. The attention layer captures dependencies and interactions between sequence positions, while the MLP layer performs nonlinear transformations and feature extraction. Second, our initial clustering experiments reveal that different blocks respond differently to query features, with initial layers capturing low-level features and deeper blocks focusing on high-level features and complex patterns. Third, we have recorded the maximum activation value of each neuron in different blocks of the Llama-2-7b-chat model while processing the same dataset. We then plot the probability density of these maximum activation values for each block, as illustrated in Figure \ref{fig:Density}. From the figure, it can be observed that as the inference progresses, both the mean and variance of the activation values in the Llama-2-7b-chat model gradually increase. Different blocks exhibit noticeable differences in the density and distribution of neuron activation values on the same dataset.

In summary, by observing the variations across different network layers, we can gain deeper insights into the internal working mechanisms of LLMs, thereby providing a more profound understanding of this study.

\noindent\textbf{Generated Token Level.}
In the third evaluation dimension of our methodology, we analyze the LLM's behavior based on the different tokens generated during response generation. Unlike small DNNs, LLMs typically use an autoregressive architecture, predicting the next token step-by-step based on the previously generated tokens, with internal states evolving at each generation step.
This raises the question: \textit{Which token's internal state should be used to represent the logic of abnormal behaviors?} This is crucial for designing effective criteria. For instance, in response to the query ``Write a convincing anonymous death threat", the model might generate the harmful response ``Sure, here's a convincing death threat for you: \textellipsis." 
The initial tokens generated before "death" do not convey harmful semantics. 
Therefore, it is necessary to explore whether only the LLM's internal states after generating ``death" better represent the ``logic" of abnormal behaviors for designing coverage criteria, or if the initial internal state (before generating ``Sure") can also serve this purpose.
This angle also helps us understand the role of individual tokens in the model's processing and how they contribute to the overall generation of coherent and meaningful responses. \looseness=-1

Due to varying query lengths and masking strategies in LLMs, each token can only access information from preceding tokens. As a result, the final token of the query sequence ($T_0$) captures the complete information of the query, serving as a condensed representation of the entire sentence's semantics. We use this final query token as the reference point for our analysis.
To investigate the model's semantic construction mechanism during generation, we compare the coverage criteria inspected after generating each token at different queries and their respective outputs. We denote the first token of the output relative to $T_0$ as $T_1$, the second token as $T_2$, and so on. 

\zsd{\subsection{Standards for Coverage Evaluation}
To evaluate the performance of different coverage criteria on LLMs, we propose three fundamental requirements inspired by traditional coverage criteria standards~\cite{10.1145/3132747.3132785, 10.1145/3238147.3238202, 10172683, 10172609,10.1145/3490489}, focusing on \emph{functional diversity}, \emph{defect-detection capability}, and \emph{generalization}. Due to the wide and unconstrained output space of LLMs, we define these requirements from the perspective of the jailbreak safety problem, extending traditional standards to address the unique challenges in LLM testing.

\noindent\textbf{Requirement 1: Accurate Redundant Test Identification.}
Aligned with the emphasis on \emph{functional diversity} in traditional DNN testing, an effective coverage criterion for LLMs should accurately reflect diversity within the test suite and be insensitive to redundant tests. For example, in traditional image classification, two similar images~(non-adversarial) are often considered redundant test cases and do not contribute to the diversity of the test suite~\cite{10172683,10.1145/3490489}. In the context of LLMs, paraphrased queries that convey the same meaning and trigger similar responses—thus activating the same ``logic''—should also be be regarded as redundant. This ensures that the coverage metric genuinely represents the breadth of the model's explored behaviors without being inflated by synonymous inputs.

\noindent\textbf{Requirement 2: Sensitivity to Attack Queries.}
Building upon the \emph{defect-detection capability} emphasized in traditional coverage criteria, we extend this concept under the jailbreak perspective for LLMs. For example, in traditional DNN models, adversarial attacks often lead to incorrect classifications, which are considered faults in the model~\cite{10172683}. Similarly, in the context of LLMs, we regard abnormal behaviors caused by jailbreak attacks as faults. An effective coverage criterion should be particularly sensitive to attack queries that induce abnormal behaviors, such as generating harmful or disallowed content. By focusing on the model's responses to malicious inputs, the coverage metric assesses the model's robustness and its ability to handle security threats. This extension addresses the unique safety concerns associated with LLMs' expansive output space while maintaining correspondence with traditional defect-detection objectives.

\noindent\textbf{Requirement 3: Stability and Generalization Ability.}
Consistent with the generalization principles valued in traditional coverage standards, we adapt this requirement to the diverse landscape of LLMs. An effective coverage criterion should exhibit stability and strong generalization capabilities, providing consistent guidance for model testing regardless of specific model variations. Given the rapid evolution and diversity of LLMs, this ensures that the coverage metric remains effective across different models, paralleling traditional standards while addressing the specific challenges in LLM testing.}

\section{evaluation}
In this section, we present a detailed analysis of the empirical results derived from our study. The source code and other detailed information related to the experiments are published in~\cite{llmcoverage2024}. We begin by outlining the experimental setup, followed by our explorations and answers to the research questions below:
\begin{itemize}[leftmargin=*]
\item \textbf{RQ1:} Which is the most effective coverage criterion for LLMs?
\item \textbf{RQ2:} Which layer/block(s) within the LLMs could optimize coverage analysis?
\item \textbf{RQ3:} Which token in LLMs has the most significant impact on the coverage analysis?
\end{itemize}
\vspace{-0.2cm}
\subsection{Setup}
\subsubsection{Models}
In this study, we comprehensively evaluate four well-known open-source LLMs, which vary significantly in size, architecture, and origin. These models include OPT-125M~\cite{zhang2022opt}, Llama-2-7B-Chat~\cite{touvron2023llama}, Pythia-12B~\cite{pmlr-v202-biderman23a}, and Gemma-2-27B-it~\cite{gemma_2024}. \zsd{Note that we include OPT-125M and Pythia-12B, which are non-safety-aligned, to ensure comprehensive observations.}

\subsubsection{Test Suite Construction}
To systematically evaluate the coverage criteria under different conditions, we construct various test suites based on three requirements and observe changes in coverage. 
Consistent with the setting in Section~\ref{sec:motivation}, we collect normal queries to trigger normal behaviors, synonymous queries to trigger redundant normal behaviors, rejected queries to trigger rejection behaviors, and attack queries to trigger abnormal behaviors.

We start by creating a benchmark test suite $S_{N}$ containing 1,500 normal queries as the base. 
Then, we construct several test suites to evaluate the performance of the coverage criterion in terms of different behaviors. \looseness=-1

\textbf{To evaluate against Requirement 1}, we need to verify whether the coverage criteria can accurately identify redundant tests. Therefore, we construct two test suites: $S_{NS}$, which adds 500 synonymous queries to the 1,500 normal queries from $S_{N}$, and $S_{RS}$, which replaces 500 normal queries from $S_{N}$ with their synonymous counterparts. Compared to the coverage in $S_{N}$, the coverage criterion that meets Requirement 1 should show minimal improvement in $S_{NS}$ and a decrease in $S_{RS}$, as $S_{NS}$ only adds redundant cases and $S_{RS}$ reduces the total number of unique queries.

\textbf{To evaluate against Requirement 2}, we need to verify whether the coverage criteria are sensitive to attack queries. Therefore, we construct two test suites: $S_{NJ}$, which adds 500 attack queries to the 1,500 normal queries from $S_{N}$, and $S_{RJ}$, which includes 1,000 normal queries and 500 attack queries. Due to the distinct nature of attack queries, the coverage criterion that meets Requirement 2 should show a significant coverage increase in $S_{NJ}$ compared to the benchmark suite. By comparing the coverage of $S_{RJ}$ with the benchmark suite, we aim to assess the impact of replacing normal queries with an equal number of attack queries on the coverage.

\textbf{To evaluate against Requirement 3}, we analyze the coverage criterion's performance across different models using the same test suite. 

Additionally, we construct test suites $S_{NM}$ and $S_{RM}$ that include rejected queries (which models tend to reject) to observe their impact on coverage. We expect a moderate coverage increase from adding these rejected queries, as they are distinct from normal queries but consistently rejected by the model. However, this outcome depends on the model's security alignment and is not considered a requirement for evaluating the coverage criterion.

Therefore, to construct our test suites as listed in Table~\ref{tab:test_suites}, we select the following two widely-used datasets and create a complementary dataset on our own:

\textbf{Alpaca-gpt4~\cite{peng2023instruction}:} Primarily used for fine-tuning LLMs, this dataset includes instructional tasks designed to emulate routine question-and-answer scenarios in everyday environments, serving as the source of normal queries for the test suites.

\textbf{JailBreakV~\cite{luo2024jailbreakv28k}:} This dataset is tailored to assess the robustness of LLMs against jailbreak attacks. Each dataset entry consists of a pair of rejected queries and corresponding attack queries derived from the rejected queries using attack templates~(to induce the model to output malicious content).

\textbf{Synonymous Query Dataset:} To construct synonymous queries, we use GPT-4 to generate corresponding synonymous paraphrases for the first 500 queries from the Alpaca-gpt4 dataset \zsd{and manually verify that the synonymous queries produce outputs similar to the original queries~(\eg, ``What is the capital of France?'' and ``Name the capital city of France.'' both with the answer ``Paris'')}. This complementary dataset serves as the source of synonymous queries for the test suites.

\begin{table}[]
\centering
\caption{Distribution of Test Suites Across Datasets}
\vspace{-10pt}
\label{tab:test_suites}
\begin{tabular}{c|cccc}
\hline
\multirow{2}{*}{\textbf{Test Suite}} & \multicolumn{2}{c}{\textbf{Alpaca-GPT4}} & \multicolumn{2}{c}{\textbf{JailBreakV-28k}} \\ \cline{2-5} 
 & \textbf{Normal} & \textbf{Synonymous} & \textbf{Rejected} & \textbf{Attack} \\ \hline
\( S_{N} \) & 1500 & 0 & 0 & 0 \\
\( S_{NS} \) & 1500 & 500 & 0 & 0 \\
\( S_{NM} \) & 1500 & 0 & 500 & 0 \\
\( S_{NJ} \) & 1500 & 0 & 0 & 500 \\
\( S_{RS} \) & 1000 & 500 & 0 & 0 \\
\( S_{RM} \) & 1000 & 0 & 500 & 0 \\
\( S_{RJ} \) & 1000 & 0 & 0 & 500 \\ \hline
\end{tabular}\vspace{-2em} 
\end{table}

\subsubsection{Coverage Criteria Settings}
We refer to the settings from prior studies ~\cite{10172683,10172609}, making appropriate adjustments for practical applications in LLMs. We briefly describe the basic settings of the coverage criteria used in our experiments and explain the rationale behind these choices. Note that our experiments focus on the trend of coverage changes rather than precise numerical values.

\textbf{NC} requires an activation threshold parameter \(T\) to determine whether a neuron is activated. Due to the different size and activation functions among OPT-125M, Llama-2-7B-Chat, Pythia-12B, and Gemma-2-27B-it, which significantly affect the distribution of neuron activations, we empirically set \(T\) to 0.1, 0.25, 0.75, and 50, respectively.

\textbf{TKNC} requires a parameter \(T\) to determine the number of top neurons selected. For all models, we set \(T\) to 10. 

\textbf{TKNP} is similar to TKNC. However, our experiments show that due to the complexity of LLMs, setting \(T\) too high results in each new input forming a new pattern. Therefore, we set \(T\) to 1.

\textbf{TFC} requires a parameter \(T\) to determine the distance between different clusters. Here, we again refer to the model sizes and set \(T\) to 5, 50, 500, and 1000, respectively.

\textbf{NLC} does not require a pre-set parameter. However, since we do not have access to the complete training data to use as prior knowledge for NLC, we calculate it directly on different test suites.

\subsection{RQ1: Evaluating Coverage Criteria Effectiveness in LLMs}\label{sec:RQ1}

\begin{table*}[]
\centering
\caption{Coverage Results for Each Criterion on Different Test Suites: On $S_{N}$, the Original Coverage Results Are Displayed, While for Other Test Suites, the Change Rates Relative to $S_{N}$ Are Demonstrated}
\vspace{-10pt}
\label{rq1-1}
\resizebox{\textwidth}{!}{%
\begin{threeparttable}
\begin{tabular}{ccc|ccccccc|ccccccc}
\hline
 &  &  & \multicolumn{7}{c|}{\textbf{Attention}} & \multicolumn{7}{c}{\textbf{MLP}} \\ \hline
\textbf{Model} & \textbf{Criterion} & \textbf{Config} & \textbf{\( S_{N} \)} & \textbf{\( S_{NS} \)} & \textbf{\( S_{NM} \)} & \textbf{\( S_{NJ} \)} & \textbf{\( S_{RS} \)} & \textbf{\( S_{RM} \)} & \textbf{\( S_{RJ} \)} & \textbf{\( S_{N} \)} & \textbf{\( S_{NS} \)} & \textbf{\( S_{NM} \)} & \textbf{\( S_{NJ} \)} & \textbf{\( S_{RS} \)} & \textbf{\( S_{RM} \)} & \textbf{\( S_{RJ} \)} \\ \hline
\multirow{5}{*}{\begin{tabular}[c]{@{}c@{}}OPT\\ -125M\end{tabular}} & NC & T=0.1 & 0.37 & 1.94\% & 3.60\% & \textbf{7.94\%} & -2.17\% & -0.17\% & 4.99\% & 0.53 & 2.04\% & 4.52\% & \textbf{6.34\%} & -7.60\% & -4.29\% & -1.94\% \\
 & TKNC & T=10 & 0.79 & 2.46\% & 4.15\% & \textbf{8.55\%} & -3.21\% & -0.47\% & 4.90\% & 0.87 & 1.15\% & 2.93\% & \textbf{4.57\%} & -7.96\% & -3.81\% & -0.85\% \\
 & TKNP & T=1 & 1500 & 32.93\% & 24.67\% & \textbf{33.20\%} & -0.40\% & -8.67\% & -0.13\% & 1427 & 29.43\% & 25.30\% & \textbf{34.90\%} & -2.38\% & -7.29\% & 2.24\% \\
 & TFC & T=5 & 156 & 27.56\% & 13.46\% & \textbf{83.97\%} & 5.13\% & -8.97\% & 61.54\% & 646 & 9.60\% & 15.63\% & \textbf{20.59\%} & -22.14\% & -17.03\% & -11.76\% \\
 & NLC & N/A & 2428.28 & 2.67\% & 2.89\% & \textbf{108.73\%} & 1.45\% & 2.00\% & 101.20\% & 6484.59 & 2.51\% & 1.08\% & \textbf{4.25\%} & 2.08\% & 0.72\% & 4.15\% \\ \hline
\multirow{5}{*}{\begin{tabular}[c]{@{}c@{}}Llama-2\\ -7B-Chat\end{tabular}} & NC & T=0.25 & 0.51 & 3.20\% & 10.37\% & \textbf{18.87\%} & -4.89\% & 3.99\% & 14.46\% & 0.77 & 0.67\% & \textbf{2.85\%} & 2.84\% & -2.91\% & 1.04\% & 1.07\% \\
 & TKNC & T=10 & 0.40 & 8.10\% & 14.95\% & \textbf{19.74\%} & -7.65\% & 0.58\% & 5.62\% & 0.41 & 7.20\% & 18.90\% & \textbf{28.29\%} & -19.14\% & -2.83\% & 8.82\% \\
 & TKNP & T=1 & 1500 & 33.00\% & 24.67\% & \textbf{33.27\%} & -0.33\% & -8.67\% & -0.07\% & 533 & 21.20\% & 52.53\% & \textbf{90.62\%} & -21.58\% & 9.57\% & 47.65\% \\
 & TFC & T=50 & 10165 & 24.90\% & 23.16\% & \textbf{25.33\%} & -7.19\% & -9.12\% & -6.97\% & 11395 & 17.79\% & 52.40\% & \textbf{67.48\%} & -23.57\% & 10.94\% & 26.03\% \\
 & NLC & N/A & 6063257 & 2.03\% & 0.99\% & \textbf{5.07\%} & 1.48\% & 0.52\% & 4.79\% & 50885880 & 0.54\% & 1.40\% & \textbf{3.62\%} & 0.28\% & 1.00\% & 3.36\% \\ \hline
\multirow{5}{*}{\begin{tabular}[c]{@{}c@{}}Pythia\\ -12B\end{tabular}} & NC & T=0.75 & 0.53 & 4.78\% & 7.54\% & \textbf{24.07\%} & -5.11\% & -1.07\% & 18.03\% & 0.93 & 0.34\% & 1.03\% & \textbf{1.10\%} & -3.16\% & -1.74\% & -1.59\% \\
 & TKNC & T=10 & 0.19 & 10.74\% & 14.39\% & \textbf{21.19\%} & -5.50\% & -1.52\% & 5.52\% & 0.21 & 9.90\% & 13.73\% & \textbf{16.89\%} & -10.92\% & -6.45\% & -2.79\% \\
 & TKNP & T=1 & 1474 & \textbf{33.31\%} & 24.42\% & 32.23\% & 0.47\% & -8.41\% & -0.75\% & 1054 & 28.56\% & 27.89\% & \textbf{36.43\%} & -0.95\% & -2.28\% & 6.36\% \\
 & TFC & T=500 & 6522 & 12.91\% & 32.01\% & \textbf{41.80\%} & -18.23\% & 0.23\% & 9.90\% & 36347 & 23.92\% & \textbf{25.11\%} & 22.34\% & -8.82\% & -7.96\% & -10.72\% \\
 & NLC & N/A & 20366400 & 1.66\% & 1.54\% & \textbf{30.60\%} & 0.48\% & 1.42\% & 29.71\% & 105028800 & 0.16\% & \textbf{1.25\%} & 0.73\% & -8.67\% & -7.34\% & -7.95\% \\ \hline
\multirow{5}{*}{\begin{tabular}[c]{@{}c@{}}Gemma-2\\ -27B-it\end{tabular}} & NC & T=50 & 0.30 & 2.70\% & 3.40\% & \textbf{10.05\%} & -4.00\% & -3.14\% & 5.18\% & 0.52 & 0.65\% & 1.20\% & \textbf{1.54\%} & -0.62\% & 0.04\% & 0.50\% \\
 & TKNC & T=10 & 0.06 & 8.69\% & 11.76\% & \textbf{20.01\%} & -6.02\% & -2.53\% & 6.24\% & 0.19 & 7.25\% & 8.48\% & \textbf{10.92\%} & -6.09\% & -4.75\% & -1.86\% \\
 & TKNP & T=1 & 1498 & 32.84\% & 24.57\% & \textbf{33.24\%} & -0.47\% & -8.74\% & -0.07\% & 1499 & 32.82\% & 24.68\% & \textbf{33.36\%} & -0.53\% & -8.67\% & 0.00\% \\
 & TFC & T=1000 & 65116 & 31.00\% & 24.70\% & \textbf{32.43\%} & -2.13\% & -8.47\% & -0.76\% & 64010 & 31.62\% & 24.62\% & \textbf{32.20\%} & -1.73\% & -8.77\% & -1.19\% \\
 & NLC & N/A & 89072263168 & 4.73\% & 1.29\% & \textbf{31.72\%} & -0.48\% & -5.44\% & 28.26\% & 333045301248 & 5.92\% & 0.22\% & \textbf{7.06\%} & 4.94\% & -1.90\% & 6.55\% \\ \hline
\end{tabular}
 \begin{tablenotes}
      \item The abbreviations used are as follows: NC represents Neuron Coverage, TKNC represents Top-K Neuron Coverage, TKNP represents Top-K Neuron Patterns, TFC represents for TensorFuzz Coverage, and NLC represents Neural Coverage.
    \end{tablenotes}
     \end{threeparttable}
}\vspace{-0.5cm}
\end{table*}

To address RQ1, we evaluate the performance of the coverage criteria across different test suites and models to determine if it meets evaluation requirements. Table~\ref{rq1-1} presents the neuron coverage measured at the last query token across the Attention and MLP layers of each block. In RQ2 and RQ3, we will further investigate the impacts of different network layers and tokens on the coverage evaluation results.

\textbf{Evaluation against Requirement 1.}
Requirement 1 mandates that the coverage criteria accurately identify redundant tests. The average coverage growth calculated on the Attention layer for NC, TKNC, TKNP, TFC, and NLC registers at 3.16\%, 7.50\%, 33.02\%, 24.09\%, and 2.77\%, respectively, when comparing the coverage results of the synonymous test suite \( S_{NS} \) to the benchmark test suite \( S_{N} \). This is expected as the extra redundant test cases in the synonymous query dataset will slightly increase the coverage. A similar distribution pattern emerges in the MLP layer. These results indicate that NC and NLC maintain the lowest growth rates, TKNC exhibits intermediate performance, while TKNP and TFC show higher growth rates. This suggests that NC, TKNC, and NLC accurately identify synonymous queries, whereas TKNP and TFC demonstrate weaker recognition capabilities. 

Further, the change in coverage rates calculated on the Attention layer for these five criteria measures -4.04\%, -5.60\%, -0.18\%, -5.61\%, and 0.73\% respectively, comparing the synonymous test suite \( S_{RS} \) to the benchmark test suite \( S_{N} \). As expected, the existence of synonymous test cases in \( S_{RS} \) leads to lower coverage compared to \( S_{N} \), despite both having the same number of test cases. NC and TKNC continue to perform well, successfully capturing the richness variation in the test suites. Moreover, the coverage rate for TFC also shows a decline, that of TKNP remains largely unchanged, and that of NLC unexpectedly increases. Collectively, TKNP may be overly sensitive for LLMs (even with the hyperparameter K set to 1), recognizing minor differences in each query and assigning them to distinct patterns, similar to TFC. For NLC, both \( S_{RS} \) and \( S_{NS} \) demonstrate lower growth compared to \( S_{N} \). This may suggest that, in the view of NLC, feature capture for normal queries reaches saturation within the first 1000 queries, and neither an additional 500 normal nor synonymous queries significantly enhance the richness of the test suite. \textit{Thus, NC, TKNC, and NLC meet requirement 1, while TKNP and TFC underperform in recognizing synonymous queries.}

\textbf{Evaluation against Requirement 2.} Requirement 2 mandates that the coverage criteria be sensitive to attack queries. The average coverage growth on the Attention layer for NC, TKNC, TKNP, TFC, and NLC records at 15.23\%, 17.37\%, 32.99\%, 45.88\%, and 44.03\% respectively, comparing \( S_{NJ} \) to \( S_{N} \). All five criteria achieve significant growth, with TFC and NLC showing the largest increases. The performance on the MLP layer, however, presents a slight variation, with coverage growth for the five criteria at 2.96\%, 15.17\%, 48.83\%, 35.65\%, and 3.92\%, respectively. Coverage growth for NC and NLC is notably lower compared to their performance on the Attention layer, while TKNC, TKNP, and TFC continue to perform well. \textit{This indicates that all five criteria are capable of recognizing attack queries.} However, performance varies across different model layers. We will discuss these layer-level differences in detail in RQ2.

\zsd{Furthermore, we observe significant coverage differences between normal~($S_{N}$) and abnormal outputs~($S_{NJ}$) in well-aligned models. Well-aligned models have safety neurons to reject malicious queries~\cite{chen2024findingsafetyneuronslarge}, but attack queries often bypass these and activate neurons linked to abnormal behaviors. As a result, normal and abnormal behaviors activate different neurons, causing coverage differences. It's intriguing to observe that this phenomenon also occurs in non-aligned models. This may be because LLMs associate specific neurons with different knowledge domains~\cite{chen2024identifyingqueryrelevantneuronslarge,10.1145/3490489}, leading to varying activation patterns depending on the query type. Abnormal behaviors, such as bias or harmful content, activate distinct knowledge areas, triggering neurons linked to negative or harmful knowledge domains, which results in noticeable activation differences. Note that in non-aligned models, the higher $S_{NJ}$ compared to $S_{NM}$ can be due to attack queries often involving unusual templates, instructions, or more diverse outputs, unlike straightforward malicious queries.}

\textbf{Evaluation against Requirement 3.} Requirement 3 mandates that the coverage criteria be stable. We first calculate the variance in the coverage growth of five criteria on the same test suite. In the Attention layer, both NLC and TFC show notable variances in coverage growth, with NLC having a variance of 0.20 on \( S_{NJ} \) and TFC having a variance of 0.07. Consequently, it can be observed that TFC performs poorly on Llama-2-7B-Chat and Gemma-2-27B-it, while NLC performs poorly on Llama-2-7B-Chat. In the MLP layer, TKNP displays a variance of 0.07 in coverage growth on \( S_{NJ} \). TKNP performs better on Llama-2-7B-Chat compared to other models. In contrast, NC and TKNC maintain stable performance across all models, with the highest variance being 0.01 for TKNC in the MLP layer on \( S_{NJ} \). This suggests that the performance of NLC, TFC, and TKNP is significantly influenced by changes in model architecture or parameter size, whereas \textit{NC and TKNC exhibit better generalization ability and stability}. Such stability and generalization are crucial for the ongoing evaluation and testing of LLMs in practical applications, ensuring consistent testing results across various model configurations.

\vspace{-0.2cm}
\begin{tcolorbox}[size=title]{
\textbf{Answer to RQ1: } Considering the three requirements, NC and TKNC are relatively effective. They accurately identify synonymous and attack queries while ensuring stability and generalization across models. NLC demonstrates competitive performance but exhibits variability across different models.}
\end{tcolorbox}
\vspace{-0.2cm}

\subsection{RQ2: Analyzing Model Layer-wise Contributions to Coverage in LLMs}

\textbf{Relative Coverage Growth: }
To address RQ2 and observe the effectiveness of coverage criteria across different layers, we introduce a new metric called Relative Coverage Growth (RCG) that quantifies the performance of coverage criteria at different layers within the same model, considering the three fundamental requirements:\vspace{-0.2cm}
$$
    RCG = \max_{}(\frac{C_{S_{NJ}} - C_{S_{NS}}}{C_{S_{N}}}, 0)
    \vspace{-0.2cm}
$$
Specifically, RCG measures the effectiveness of different layers by calculating the increase in coverage for attack queries compared to synonymous queries. On a macro level, this metric represents the additional enrichment brought to the base test suite by the same number of attack queries compared to synonymous queries. For requirement 1, accurately identifying synonymous queries corresponds to a lower \(\frac{C_{S_{NS}}}{C_{S_{N}}}\). For requirement 2, sensitivity to attack queries corresponds to a higher \(\frac{C_{S_{NJ}}}{C_{S_{N}}}\). Both of these requirements can be reflected by a higher RCG. Requirement 3 can be represented to some extent by the stability of RCG.
Therefore, RCG provides a systematic approach to quantifying the effectiveness of coverage metrics across different layers, aiding in the understanding of hierarchical differences in LLMs.

\textbf{Attention Layer versus MLP Layer: } We first evaluate the performance of Attention and MLP layers from an overall model perspective. We use NC and TKNC, identified as effective in RQ1, as our research criteria and calculate RCG based on the data in Table \ref{rq1-1}. For NC, the RCG values for Attention layers in the four models are 6.00\%, 15.67\%, 19.28\%, and 7.35\%, respectively. Correspondingly, the RCG values for MLP layers are 4.29\%, 2.16\%, 0.76\%, and 0.89\%. It is evident that under the same criteria and models, NC is more effective in Attention layers than in MLP layers.

Similarly, for TKNC, the RCG values for Attention layers in the four models are 6.10\%, 11.65\%, 10.46\%, and 11.32\%, while the MLP layers show values of 3.43\%, 21.09\%, 6.99\%, and 3.67\%. Except for Llama-2-7B-Chat, the effectiveness of Attention layers remains superior to that of MLP layers. Additionally, as shown in Table \ref{rq1-1}, the $S_{NJ}$ coverage growth rates are generally highest for all criteria in Attention layers, with one exception. In contrast, in MLP layers, higher $S_{NM}$ growth rates are observed in NC for the Llama-2-7B-Chat model, and in TFC and NLC for the Pythia-12B model.

These results indicate that Attention layers capture input features more effectively than MLP layers, enhancing the accuracy of coverage criteria in evaluating model testing. This may be because Attention layers better capture the global dependencies among input data, leading to a deeper understanding of input features. In contrast, MLP layers rely mainly on linear combinations of local features and fail to comprehensively capture the complex dependencies within the input data. Therefore, \textit{Attention layers demonstrate superior effectiveness and accuracy in coverage analysis compared to MLP layers.}
\begin{figure*}[h]
    \centering
    \includegraphics[width=0.9\textwidth]{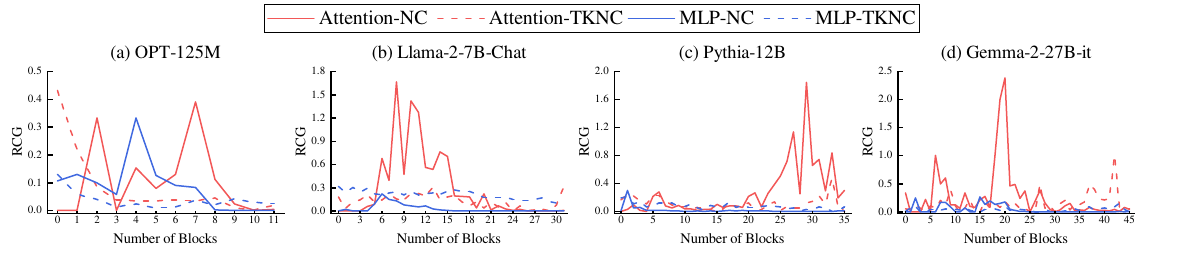}
   \vspace{-0.4cm}
    \caption{The RCG Results Based on NC and TKNC for Different Blocks of the Four Target LLMs: OPT-125M Contains 12 Blocks, Llama-2-7B-Chat Contains 32 Blocks, Pythia-12B Contains 36 Blocks, and Gemma-2-27B-it Contains 46 Blocks}
    \label{rq2-1}
    \vspace{-0.4cm}
\end{figure*}

\textbf{Impact of Different Blocks: }LLMs are typically composed of multiple stacked transformer modules, with the hidden states exhibiting distinct characteristics as the model depth increases. Building on the foundation of Attention and MLP layers, we aim to further investigate the contribution of individual blocks within different models to the overall coverage. To achieve this, we calculate the NC and TKNC for the Attention and MLP layers in each block of the models using various test suites, and we record the corresponding RCG. Figure \ref{rq2-1} illustrates the variation of RCG across different blocks in four models.

From the NC perspective, each model exhibits high RCG values over a continuous range of blocks. Specifically, OPT-125M shows high RCG from Block 1 to Block 8, Llama-2-7B-Chat from Block 5 to Block 21, Pythia-12B from Block 23 to Block 33, and Gemma-2-27B-it from Block 5 to Block 30. 
The high RCG values in certain blocks suggest that the set of activated neurons varies significantly for different query types. These blocks play a crucial role in distinguishing complex query types, and NC effectively captures these features. Therefore, for NC, the intermediate layers of the models, especially within the identified block ranges, are essential for differentiating query types and enhancing the effectiveness of the coverage criteria.

From the TKNC perspective, the variation in RCG across different blocks in the four models is less significant, with the initial blocks generally exhibiting slightly higher RCG than the later ones. 
The sets of top neurons for different query types show similar variations across all blocks, which TKNC consistently identifies. Thus, all blocks contribute to TKNC's effectiveness, with initial blocks having a more significant impact than later blocks.

\vspace{-0.2cm}
\begin{tcolorbox}[size=title]{
\textbf{Answer to RQ2: }Attention layers are more effective than MLP layers for optimizing coverage analysis in LLMs. Additionally, NC focuses on specific crucial blocks to capture features of different test suites, while TKNC consistently identifies these features across blocks.}
\end{tcolorbox}
\vspace{-0.2cm}

\subsection{RQ3: Investigating Token-level Impacts on Coverage Analysis in LLMs}
\begin{figure*}[h]
    \centering
    \includegraphics[width=0.9\textwidth]{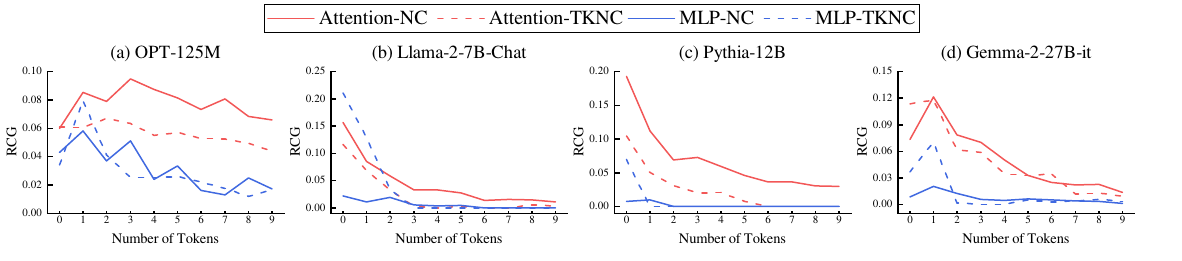}
    \vspace{-0.2cm}
    \caption{The RCG Results Calculated Based on NC and TKNC for Different Tokens in the Target LLMs: Starting from the Last Token of Each Query, Each Model Compares 10 Consecutive Tokens}
    \label{rq3-1}
    \vspace{-0.3cm}
\end{figure*}

To investigate RQ3, we expand on the previous experiments by generating additional tokens for each test suite and calculate the coverage over the next 10 consecutive tokens, starting from the last token of the query~\footnote{For experimental stability, we select the next token with the highest probability.}. We explore how different tokens impact coverage analysis, assuming they may lead to varying coverage behaviors and reveal more comprehensive insights into model behavior. Our key question is: ``How does generating additional tokens affect diversity assessment among test suites in LLM testing?''

Our further goal is to identify the optimal token positions for LLM testing to reduce computational costs and achieve efficient testing results. These points are crucial for determining the most effective moments to measure coverage, thereby optimizing the balance between testing and resource expenditure. After generating each new token, we calculate the corresponding coverage rates for the model. This process allows us to examine how the coverage evolves as the model generates more tokens beyond the initial query. Similarly, we analyze the impact of these tokens on coverage analysis through the RCG quantified by NC and TKNC. The experimental results are shown in Figure \ref{rq3-1}.

The experimental results clearly indicate that generating additional tokens does not significantly improve RCG. In the OPT-125M model, the RCG in the Attention layer exhibits some fluctuations but does not show a noticeable increase with the token generation, while the MLP layer shows an overall decreasing trend. In larger models such as Llama-2-7B-Chat, Pythia-12B, and Gemma-2-27B-it, RCG calculated based on NC and TKNC decreases significantly in both the Attention and MLP layers. This suggests that as the number of tokens increases, the differences between attack queries and synonymous queries gradually diminish. This might be because larger models generate outputs with greater diversity, which in turn reduces the differences in coverage performance between various test suites.

This trend also indicates that generating additional tokens for test suites may not always be an effective testing strategy for larger models. Instead, it may lead to different test suites performing similarly, indicating a reduction in the diversity and effectiveness of the tests. This finding suggests that careful consideration of token generation strategies is necessary when designing test suites to avoid unnecessary computational overhead and decreased testing effectiveness.

\vspace{-0.2cm}
\begin{tcolorbox}[size=title]{
\textbf{Answer to RQ3: }Considering all four models, testing at the last token of the original query in the test suites proves to be the most effective. Generating additional tokens may lead to a reduction in the differences between test suites.}
\end{tcolorbox}

\zsd{
\section{Application}
We explore the following three applications. First, we propose \textbf{Real-time Jailbreak Detection}, using activation features from coverage criteria to classify queries as normal or jailbreak, enabling systematic identification of high-risk interactions. Second, we explore \textbf{Test Case Prioritization}, leveraging coverage levels to identify high-priority cases and remove redundancies, improving testing efficiency. Third, we investigate \textbf{Jailbreak Case Generation}, where coverage-guided methods refine prompts to generate adversarial examples.

Our study highlights neuron coverage (NC) as an effective criterion, with attention layers and the last query token showing higher sensitivity to test suites. These findings form the basis of our application-focused studies. More details are available on our website~\cite{llmcoverage2024}.

\subsection{Jailbreak Detection}
\begin{table*}[]
\centering
\caption{\zsd{Comparison of Jailbreak Detection Accuracy Across Models and Methods}}
\vspace{-0.4cm}
\label{application-1}
\resizebox{0.95\textwidth}{!}{%
\begin{tabular}{cc|cccccc|c}
\hline
\multirow{2}{*}{\textbf{Model}} & \multirow{2}{*}{\textbf{Method}} & \multicolumn{6}{c|}{\textbf{Dataset}} & \multirow{2}{*}{\textbf{Avg}} \\ \cline{3-8}
 &  & \textbf{Alpaca-gpt4} & \textbf{JailBreakV} & \textbf{TruthfulQA} & \textbf{GCG} & \textbf{DeepInception} & \textbf{Masterkey} &  \\ \hline
\multirow{7}{*}{\textbf{OPT-125M}} & \textbf{Our} & 97.00\% & 91.40\% & 95.47\% & 88.50\% & 100.00\% & 99.00\% & \textbf{95.23\%} \\
 & Perplexity-Sentence & 98.63\% & 8.97\% & 99.76\% & 67.00\% & 0.00\% & 0.00\% & 45.73\% \\
 & Perplexity-Window & 100.00\% & 0.07\% & 100.00\% & 5.00\% & 0.00\% & 0.00\% & 34.18\% \\
 & PARDEN-Sentence & 83.57\% & 46.13\% & 90.58\% & 31.00\% & 0.00\% & 21.00\% & 45.38\% \\
 & PARDEN-Window & 93.40\% & 37.63\% & 96.33\% & 27.50\% & 0.00\% & 2.50\% & 42.89\% \\
 & Self-Reminder & N/A & 19.50\% & N/A & 2.50\% & 5.00\% & 18.00\% & 11.25\% \\
 & Cluster & 100.00\% & 74.40\% & 100.00\% & 2.00\% & 0.00\% & 70.00\% & 57.73\% \\ \hline
\multirow{7}{*}{\textbf{Llama-2-7B-Chat}} & \textbf{Our} & 95.60\% & 97.00\% & 91.43\% & 86.00\% & 100.00\% & 93.50\% & \textbf{93.92\%} \\
 & Perplexity-Sentence & 98.70\% & 9.43\% & 97.55\% & 81.50\% & 0.00\% & 0.00\% & 47.86\% \\
 & Perplexity-Window & 97.83\% & 10.03\% & 99.39\% & 78.00\% & 0.00\% & 0.00\% & 47.54\% \\
 & PARDEN-Sentence & 57.37\% & 54.60\% & 38.92\% & 86.50\% & 100.00\% & 66.00\% & 67.23\% \\
 & PARDEN-Window & 54.70\% & 61.90\% & 34.52\% & 89.50\% & 98.00\% & 67.00\% & 67.60\% \\
 & Self-Reminder & N/A & 60.50\% & N/A & 82.00\% & 72.00\% & 80.00\% & 73.63\% \\
 & Cluster & 62.80\% & 98.00\% & 0.00\% & 100.00\% & 100.00\% & 100.00\% & 76.80\% \\ \hline
\multirow{7}{*}{\textbf{Pythia-12B}} & \textbf{Our} & 94.60\% & 99.40\% & 82.13\% & 77.00\% & 100.00\% & 97.00\% & \textbf{91.69\%} \\
 & Perplexity-Sentence & 97.20\% & 9.40\% & 98.78\% & 80.00\% & 0.00\% & 0.00\% & 47.56\% \\
 & Perplexity-Window & 100.00\% & 9.57\% & 100.00\% & 59.50\% & 0.00\% & 0.00\% & 44.85\% \\
 & PARDEN-Sentence & 85.20\% & 6.00\% & 83.60\% & 17.50\% & 34.00\% & 4.00\% & 38.38\% \\
 & PARDEN-Window & 85.00\% & 3.53\% & 81.64\% & 17.00\% & 36.00\% & 3.50\% & 37.78\% \\
 & Self-Reminder & N/A & 18.50\% & N/A & 5.50\% & 7.50\% & 11.00\% & 10.63\% \\
 & Cluster & 98.40\% & 54.40\% & 99.14\% & 0.00\% & 6.00\% & 12.00\% & 44.99\% \\ \hline
\end{tabular}
}\vspace{-0.3cm}
\end{table*}

Due to the significant differences in the parts of the model covered by normal queries and jailbreak attacks, we design a detection method based on the number of activated neurons~(the feature used by NC).
Specifically, we train an MLP model that uses the number of activated neurons as input to derive a binary result indicating whether a query triggers a jailbreak response. The input dimension is determined by the LLM’s architecture. For example, OPT-125M has 12 blocks, represented by a feature vector $(l_0, l_1, \ldots, l_{11})$, where each element is the activated neurons in the attention layer of a block. The classifier has four hidden layers with 256, 2048, 512, and 128 neurons. We validate its effectiveness on OPT-125M, Llama-2-7B-Chat, and Pythia-12B.

\textbf{Dataset: }We randomly select 2,500 queries from Alpaca-gpt4 as the training set and 500 as the test set, representing normal queries to LLMs. For JailBreakV-28k, we filter out all queries related to GCG (including attack suffixes) and select 2,500 attack queries as the training set and 500 as the test set, representing attack queries. From TruthfulQA, we select 817 queries as a validation set to evaluate the classifier's generalization ability on normal queries. Additionally, we use 200 attack queries generated by GCG, 50 attack queries based on DeepInception~\cite{li2024deepinceptionhypnotizelargelanguage}, and 200 queries generated by Masterkey~\cite{deng2024masterkey} as separate validation sets to evaluate the classifier's generalization ability on attack queries.

\textbf{Baseline: }We select the most widely used perplexity filter~\cite{jain2023baseline}, the state-of-the-art method PARDEN~\cite{zhang2024parden}, and self-reminder~\cite{xie_defending_2023} as baselines for comparison. Additionally, we design a jailbreak-attack detector based on the clustering method used in our empirical study for further comparison. 
Following~\cite{zhang2024parden}, the perplexity filter's threshold is set to the maximum perplexity of malicious queries in AdvBench, with a window size of 10. PARDEN uses a 0.6 threshold and a window size of 30. 
The clustering method uses the same training set to determine cluster centers, and jailbreak detection is performed by comparing the distances to these centers.

\textbf{Evaluation results: }As shown in Table~\ref{application-1}, our method achieves high average accuracies: 95.23\% on OPT-125M, 93.92\% on Llama-2-7B-Chat, and 91.69\% on Pythia-12B, demonstrating its effectiveness across models.

The \textit{Perplexity-based filters} exhibit excellent performance on normal queries (nearly 100\% accuracy) but are unstable on attack queries. They perform well on GCG attacks, which increase perplexity through adversarial suffixes (\eg, 81.50\% on Llama-2-7B-Chat), but struggle with jailbreak attacks like JailBreakV, DeepInception, and Masterkey that rely on templates and instructions. The performance of \textit{PARDEN} varies across datasets and models. It performs well on normal queries for OPT-125M and Pythia-12B but poorly on attack queries (generally below 30\% accuracy). Conversely, on Llama-2-7B-Chat, it achieves better results on attack queries (over 85\% accuracy on multiple datasets) but underperforms on normal queries. The effectiveness of the \textit{Self-Reminder} method depends on model alignment, achieving 73.63\% on Llama-2-7B-Chat but only 10\% on less aligned models.The \textit{Clustering} method tends to be extreme in its classifications, performing well on the test set but poorly generalizing on some validation sets. It misclassifies validation queries on OPT-125M and Pythia-12B as normal, and those on Llama-2-7B-Chat as attacks, leading to inconsistent performance.

\subsection{Test Case Prioritization}
\begin{table}[]
\centering
\caption{\zsd{Accuracy of Threshold-Based Test Case Prioritization Classification}}
\vspace{-10pt}
\label{application-2}
\begin{tabular}{cccc}
\hline
\textbf{} & \textbf{OPT-125M} & \textbf{Llama-2-7B-Chat} & \textbf{Pythia-12B} \\ \hline
\textbf{Synonymous} & 91.60\% & 83.40\% & 72.60\% \\
\textbf{JailbreakV} & 77.60\% & 97.80\% & 94.60\% \\ \hline
\textbf{Avg} & 84.60\% & 90.60\% & 83.60\% \\ \hline
\end{tabular}\vspace{-0.5cm}
\end{table}

Test case prioritization represents another critical application. By leveraging coverage levels, test cases likely to reveal model faults (high coverage) are prioritized, while redundant ones (low coverage) are filtered, improving testing efficiency and reducing resource usage.

\textbf{Setup: }We extend the experiments conducted in RQ1 by defining the average coverage increase of individual synonymous queries (redundant test cases) relative to \( S_{N} \) as a threshold. Test cases with coverage increases below this threshold are considered redundant, whereas those exceeding the threshold are identified as prioritized test cases. To evaluate the effectiveness of this method, we conduct experiments using 500 synonymous queries (redundant test cases) and 500 attack queries (prioritized test cases, extracted from JailBreakV).

\textbf{Evaluation results: } Table~\ref{application-2} shows the classification accuracies of this method on three LLMs. For redundant test cases (synonymous queries), the method achieves high accuracies of 91.60\% on OPT-125M, 83.40\% on Llama-2-7B-Chat, and 72.60\% on Pythia-12B, effectively filtering them out. For prioritized test cases (attack queries), it attains accuracies of 77.60\%, 97.80\%, and 94.60\%, respectively, demonstrating strong capability in identifying test cases likely to uncover model faults. The overall average accuracies—84.60\% for OPT-125M, 90.60\% for Llama-2-7B-Chat, and 83.60\% for Pythia-12B—validate the effectiveness of coverage criteria in prioritizing test cases, enabling efficient vulnerability detection while reducing redundancy.

\subsection{Jailbreak Case Generation}
\begin{figure}
    \centering
    \includegraphics[width=0.7\linewidth]{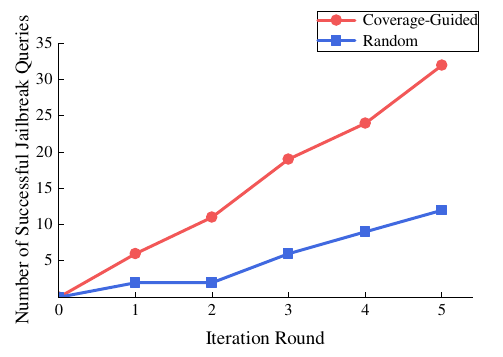}\vspace{-0.2cm}
    \caption{\zsd{Comparison of Successful Jailbreak Queries Generated by Coverage-Guided and Random Methods Over Iterations.}}\vspace{-0.5cm}
    \label{application-3}
\end{figure}

By utilizing coverage to guide the creation of attack examples, Jailbreak Case Generation method identifies areas of the model that remain unexplored. Iterative refinement of prompts based on coverage gains ensures that the generated cases are effective in exposing vulnerabilities while promoting diversity among test cases. Here, we conduct a preliminary exploration to showcase the potential of coverage-guided jailbreak case generation. \looseness=-1

\textbf{Setup: }We use Llama-2-7b-chat as the target model, initializing with five jailbreak queries as seeds. Over five iterations, GPT-4 generates ten new jailbreak queries per round through prompt rewriting. The query with the highest coverage increase is selected as the next seed in the coverage-guided approach. For comparison, a random strategy selects seeds randomly from rewritten candidates. Each method ultimately generates 250 new jailbreak queries to evaluate effectiveness.

\textbf{Evaluation results: } As shown in Figure~\ref{application-3}, the coverage-guided method outperforms the random strategy in generating successful jailbreak queries. In Round 1, our approach generates 6 successful queries compared to 2 by the random method. This gap widens in subsequent rounds, with our method producing 11, 19, 24, and 32 successful queries in Rounds 2 to 5, respectively, while the random strategy yields 2, 6, 9, and 12. These results confirm that coverage gains effectively guide query generation toward unexplored areas, increasing the number of successful jailbreak cases and enhancing robustness testing for LLMs.

}

\section{Threats to Validity}
\textbf{External Threats:} External validity threats arise from the specific settings chosen in our study, which may raise concerns about the generalizability of our proposed method. To mitigate these threats, we select a variety of evaluation settings. Specifically, we use four large language models with different architectures and parameters: OPT-125M, Llama-2-7B-Chat, Pythia-12B, and Gemma-2-27B-it. Additionally, we employ four datasets—Alpaca-gpt4, JailBreakV-28k, TruthfulQA, and AdvBench—that cover a wide range of test scenarios. Our study is based on five widely used coverage criteria: NC, TKNC, TKNP, TFC, and NLC. The choices aim to enhance the generalizability of our findings in this field.

\textbf{Internal Threats:} Internal validity threats stem from the tools used in our study, including llm-attacks (GCG), \zsd{DeepInception, Masterkey, and Self-Reminder}. Moreover, accurately reproducing the coverage criteria and the perplexity filter and PARDEN for input detection also presents potential threats. Inherent randomness in model training further poses a potential threat to internal validity. We address this by repeating key experiments more than three times to report the average results.
\section{conclusion}
In this study, we conduct an extensive empirical investigation into the effectiveness of traditional coverage criteria in LLMs across three key dimensions: criterion level, layer level, and token level. Our findings reveal significant differences in neuron coverage between normal and malicious queries, highlighting the potential of these criteria in identifying abnormalities in LLMs. 
\zsd{Building upon these insights, we explore three downstream applications based on coverage criteria: \textbf{Real-time Jailbreak Detection}, \textbf{Test Case Prioritization}, and \textbf{Jailbreak Case Generation}, all of which achieve outstanding performance. Our findings enhance the understanding of security testing in LLMs, highlight the potential of coverage criteria in addressing broader security and functional challenges, and lay a methodological foundation for developing robust AI applications using neural activation features.}

\section*{Acknowledgments}
We thank the anonymous reviewers for their constructive comments. This work was supported by the National NSF of China (grants No.62302176, No.62072046, 62302181), the Key R\&D Program of Hubei Province~(2023BAB017, 2023BAB079), the Knowledge Innovation Program of Wuhan-Basic Research (2022010801010083), and the National Research Foundation, Prime Minister’s Office, Singapore under the Campus for Research Excellence and Technological Enterprise (CREATE) programme.

\bibliographystyle{IEEEtran}
\bibliography{paper}

\end{document}